# Comparison between the diffuse interface and volume of fluid methods for simulating two-phase flows

Shahab Mirjalili[a,*], Christopher B. Ivey[b], Ali Mani[a]

[a]*Department of Mechanical Engineering, Stanford University, Stanford, CA 94305, USA*
[b]*Cascade Technologies, Inc., Palo Alto, CA 94303, USA*

## Abstract

A wide variety of interface capturing methods have been introduced for simulating two-phase flows throughout the years. However, there is a noticeable dearth of literature focusing on objective comparisons between these methods, especially when they are coupled to the momentum equation and applied in physically relevant regimes. In this article, we compare two techniques for simulating two-phase flows that possess attractive qualities, but belong to the two distinct classes of diffuse interface (DI) and volume of fluid (VOF) methods. Both of these methods allow for mass-conserving schemes that can naturally capture large interfacial topology changes omnipresent in realistic two phase flows. The DI solver used in this work is based on a conservative and bounded phase field method, developed recently. Similar to level set methods, this diffuse interface method takes advantage of the smoothness of the phase field in computing curvature and surface tension forces. Geometric VOF methods track the fractional tagged volume in a cell. The specific geometric VOF scheme used here is a discretely conservative and bounded implementation that uses geometric algorithms for unsplit advection and interface reconstruction, while employing height functions for normal and curvature calculation. We present a quantitative comparison of these methods on Cartesian meshes in terms of their accuracy, convergence rate, and computational cost using canonical two-dimensional (2D) two-phase test cases: a very dense drop moving through a quiescent gas, the Rayleigh-Taylor instability, an equilibrium static drop, an oscillating drop and the damped surface wave. We further compare these methods in their ability to resolve thin films by simulating the impact of a water drop on a deep water pool. Using results of these studies, we suggest qualitative guidelines for selection of schemes for two-phase flow calculations.

## 1. Introduction

In general, numerical methods for simulating immiscible two phase flows can be split into front tracking and front capturing methods [1]. For example, in the realm of front tracking methods, Unverdi and Tryggvason [2] coupled the Navier-Stokes equations with the use of marker points for following the immiscible fluid interfaces. In these methods, marker points are advected and marker functions are reconstructed from their location. We refer the reader to a more comprehensive overview of such methods provided in [3]. These methods are not easy to adopt for capturing many realistic flows where severe topological changes can occur at the interfaces. The most popular methods among the front capturing schemes, which require no additional treatment to capture topological changes, are categorized into volume of fluid methods, level set methods and phase field (diffuse interface) methods. The interested reader is referred to [1] for a detailed review of interface capturing methods.

*Corresponding author: Phone: 650-561-5951
  *Email addresses:* `ssmirjal@stanford.edu` (Shahab Mirjalili), `civey@cascadetechnologies.com` (Christopher B. Ivey), `alimani@stanford.edu` (Ali Mani)

Volume of fluid (VOF) methods are the oldest and most commonly used class of two-phase CFD methods to this day [4, 5, 6]. These methods can be categorized based on their advection and reconstruction routines [7]. While algebraic VOF methods are simple and do not require tedious tracking of intersections, geometric VOF methods have become more widespread due to their higher accuracy. The piecewise-linear interface calculation (PLIC) VOF schemes have become particularly popular [8] because they can leverage geometric and analytic routines for the advection and reconstruction steps. This allows for the development of efficient toolboxes for performing these operations in an accurate and conservative manner. The advection operation for PLIC-VOF can be either directionally split or unsplit. Directionally split advection routines are simpler to implement, but incur extra cost due to their strict time-step requirement and their need for separate advection and reconstruction operations for each dimension. Moreover, split advection schemes are not suitable for unstructured meshes, which are critical for realistic two-phase systems. Unsplit methods on the other hand, require only one advection and reconstruction step per time-step [9, 10]. It was shown in [11] that unsplit VOF advection schemes are more accurate than their split counterparts when handling sharp interfacial features. In addition, unsplit methods are more suitable for unstructured meshes, evidenced by the development of accurate unsplit advection schemes for arbitrary non-convex polyhedral meshes [12, 13]. On the downside, unsplit schemes require expensive computation of complex flux polyhedra in interfacial cells. This can potentially lead to difficulties with parallel load balancing [14]. The computations during the reconstruction step of PLIC-VOF methods are mostly spent on accurate estimation of normal vectors after volume fraction advection is completed. The discontinuous nature of the volume fraction field presents a challenge for calculation of the normal vectors. Thus, several methods for calculation of normal vectors (and interface reconstruction based on them) have been proposed. While many schemes have emerged [15, 16, 11, 17, 18], these methods either do not offer second order accuracy or achieve that through iterative costly geometric operations. A well-known non-iterative method for estimating the normal vectors is the height-function (HF) technique [19, 20]. This method has received significant attention recently. In [21], a framework was presented for extending the HF method to 3D unstructured non-convex polyhedral meshes with application to PLIC-VOF. Their proposed method, which uses embedded height-functions (EHF) calculates interface normals with second order accuracy, with much less cost than other existing second order accurate methods.

Level set methods were introduced by [22] and further developed by many other authors. The idea of using level set methods for capturing interfaces in two-phase flows was introduced by [23]. Level set methods use the signed distance function to define an interface location. In a subsequent interface advection process, the level set field diverges from the signed distance function, potentially deteriorating the accuracy of the representation. To maintain accuracy, the values of the level set function are periodically reinitialized to represent signed distances but with minimal intrusion to the physical interface location. Due to the smoothness of the field, standard numerical schemes can be applied in the advection step of these methods and they can naturally handle topological changes. Moreover, since the level set function (a signed distance function) is a smoothly varying field, it can be conveniently used to obtain the normal and curvature values for the calculation of surface tension forces. While level set methods continue to be popular for two-phase flow simulations, inherent lack of mass conservation can be problematic for simulations performed on coarse meshes, two-phase flows with small features, and simulations with long time horizons. Improvements by [24] and [25] have reduced the mass loss by introducing a refined auxilliary mesh and combining with a PLIC-VOF method, respectively. Nevertheless, these methods introduce complexities and extra costs that the original level set methods did not have. Another solution was given by [26] and [27], where they presented the conservative level set approach. This method conserves the mass of each phase and preserves most of the desirable properties of the original level set method. Converse to VOF methods and traditional level set methods, conservative level set (CLS) is a diffuse interface method where the interface acquires a hyperbolic tangent profile in the interface-normal direction. Advection of the interface distorts this profile shape and reinitialization is required to correct the deviations and restore the desired shape of the interface after any advection step.

Phase field methods are a class of interface capturing schemes that have recently become popular for simulating immiscible two-phase flows [28, 29, 30, 31, 32, 33, 34]. These methods which are also often known as diffuse interface methods do not require any reinitialization step, unlike CLS methods. In other words, one only needs to time-integrate the phase field and momentum equations at each time-step of these methods. Typically, the Allen-Cahn or Cahn-Hilliard [35] equations are used in phase field methods. These methods admit discrete energy laws that guarantee boundedness of total energy in the domain as explained in [33] and [36]. While the Allen-Cahn method is not particularly suitable for two-phase flow simulations due to a lack of mass conservation, the Cahn-Hilliard equation is mass-conserving. However, the fourth-order derivative in the Cahn-Hilliard phase field equation incurs numerical difficulties. Furthermore, even at the PDE level, the Cahn-Hilliard equation produces shifts in the value of the phase function far from the interface, making it rather unsuitable for large density ratios [37]. Recent works have therefore sought to combine Allen-Cahn and Cahn-Hilliard equations in new phase field models that possess the advantages of both equations but avoid their issues [31, 38]. The nonconservative phase field model proposed in [31], and the CLS method inspired the conservative second order phase field method of [38]. Their method conserved the total mass similar to CahnHilliard equation, while being simpler to implement. However, in spite of a dispersion-relation-preserving upwind scheme developed for the convection-diffusion equation, the discretized phase field variable would not remain bounded between the values for the two pure phases. Numerical overshoots and undeershoots were handled via unphysical mass-clipping and mass-redistribution algorithms. In [39], a space-time discretization for the phase field equation of [38] was introduced for staggered grids. It was analytically and numerically shown that with appropriate selection of the free parameters in the phase field equation, time-integration of the phase field would always guarantee bounded values for the phase field. The resulting phase field (diffuse interface) method avoids diffusive upwinding, unphysical mass redistribution and interface reinitialization.

Before specifying the two interface capturing methods we have chosen to compare, an overview of surface tension force implementations is due. As described in [40], surface tension forces can be either implemented as stresses or body forces, respectively known as integral and volumetric formulations. While integral formulations conserve the momentum of the system automatically, no version of this type of formulation is yet to be found that allows for the discrete balance of pressure gradients and surface tension forces. This imbalance leads to the unwelcome creation of spurious currents. Volumetric formulations on the other hand, are amenable to balanced force methods that eliminate spurious currents if curvature is computed exactly [18]. Hence, volumetric formulations are adopted for modeling capillary forces in most two-phase solvers. For all such methods, calculation of normal vectors and subsequently curvature values is an important step that can significantly affect accuracy of solutions. Curvature of an interface can be defined (and computed) as the rate that the normal vector turns along the interface. As a result, accurate calculation of the normal vectors is required for accurate prediction of curvatures and surface tension forces. Two of the most popular methods for implementation of surface tension forces in VOF solvers are the convolved VOF [41] and reconstructed-distance function [41, 42] methods. Such methods however do not converge under mesh refinement. Alternatively, height functions can be employed for computing curvatures [41]. The height function (HF) method achieves second order convergence on uniform Cartesian meshes [43, 44, 45]. In [20], a second order HF technique was presented for quad and octree discretizations. For general 3D unstructured non-convex polyhedral meshes, the embedded height function (EHF) technique by [21] provided the first convergent method for computing curvatures on unstructured grids. Curvature evaluation is much more straight-forward in level set and phase field methods. In such methods, a smooth field is available from which the computation of normal vectors and curvature is possible via spatial differentiation. After computing normal vectors and curvatures the surface tension force is often implemented as a continuum surface force [42] or by way of ghost fluid method [46].

In this work, we compare the accuracy and computational cost of two recently developed VOF and DI two-phase solvers using a variety of canonical problems. In particular, our tests assess the performance of the fully coupled two phase flow solvers, where at each time-step in addition to the interface transport step, the coupled momentum equation is solved. The first solver uses a finite volume PLIC-VOF method with unsplit advection [13]. This approach is second-order accurate and discretely conservative and bounded on general non-convex polyhedral meshes. For surface tension force calculation in this method, the embedded height function (EHF) of [21] is used. The second solver uses a finite difference phase field method presented in [39]. The phase transport equation is coupled to Navier-Stokes and the continuum surface force (CSF) method is implemented for surface tension forces. More details on these solvers is provided in Section 2.

Although many numerical recipes have been developed for handling realistic two-phase flow simulations, there is a lack of understanding with regard to the relative cost-accuracy trade-off of all these methods. Comparisons of different components of two-phase flow solvers are common. For example, interface capturing schemes have been compared via standard advection tests, with prescribed velocity fields [11, 47, 48, 49]. Moreover, surface tension implementations have been compared in articles such as [41] and [50]. However, studies like [51] and [52], where the relative performance of two full two-phase solvers are compared, are scarce. In [51], the number of test cases were limited and the selected methods were quite similar. In [52], by using a common Navier-Stokes solver, 3D tests were used to compare the performance of four different interface capturing schemes with similar curvature calculation methods utilizing level sets. In their study, however, all four schemes tested were VOF methods, and because of the introduction of level set all schemes involved some mass conservation errors. The selected DI and VOF methods considered in the present study are both mass-conservative interface capturing schemes. Furthermore, they are on opposite ends of the spectrum in terms of cost and implementation complications.

Although both of these methods are bounded, mass conserving front-capturing schemes that are capable of automatically representing large interfacial changes, some clear differences between the two approaches should be emphasized. The VOF method is a sharp interface approach. On the other hand, the DI technique thickens a realistic phase transition zone from a few nanometers to few cells width. Figure 1 depicts the representation of a drop with diameter 0.5 in a $1 \times 1$ domain using the DI and VOF methods at two resolutions. It is clear that the interface location is approximately captured in a cell using discontinuous planes in the VOF method, while the DI method smears the interface over a few cells. Figure 1 gives the impression that VOF is more accurate than DI. Indeed, we will show that at the same resolution, VOF is more accurate than DI. On the other hand, VOF is much more expensive than DI. In this work the central theme is to clarify that for different types of problems which of the two methods is more accurate at equal cost levels.

In what follows, we first introduce the two specific DI and VOF numerical schemes chosen for comparison in Section 2. Section 3 compares the accuracy and cost of the these numerical methods on several canonical two-phase problems. From the results of these studies, we make recommendations for using these methods to simulate two-phase flows. Section 4 concludes the article with a summary of our findings. Appendix A presents classical curvature and advection tests for the DI method. For details on the VOF method, we refer the interested reader to [53].

## 2. Numerical Methods

### 2.1. Volume of Fluid Method

Piecewise-linear interface calculation (PLIC) VOF methods describe the interface by a series of disconnected planes, each oriented by a unit normal, $\hat{n}$, and positioned by a constant, $C$, such that

$$\vec{n} \cdot \vec{x} + C = 0, \tag{1}$$

where $\vec{n}$ points outward with respect to the reference phase, and $C$ enforces the volume fraction in the cell [54]. PLIC-VOF methods have the potential to provide discrete mass conservation, which is important to many two-phase engineering applications including those involving large density ratios, such as breaking waves, and those involving wide range of length scales associated with secondary breakups to disperse phases. In our adopted methodology, the equations governing the PLIC-VOF scheme are as follows. The volume fraction, which is advected geometrically using the non-intersecting flux polyhedron advection (NIFPA) scheme described in [55] is governed by:

$$\frac{\partial f}{\partial t} + \nabla \cdot (f\vec{u}) = 0 \tag{2}$$

The momentum equation is given by

$$\frac{\partial \rho \vec{u}}{\partial t} + \nabla \cdot (\rho \vec{u} \otimes \vec{u}) = -\nabla P + \nabla \cdot \left( \mu \left( \nabla \vec{u} + \nabla \vec{u}^\top \right) \right) + \rho \vec{g} + \vec{T}, \tag{3}$$

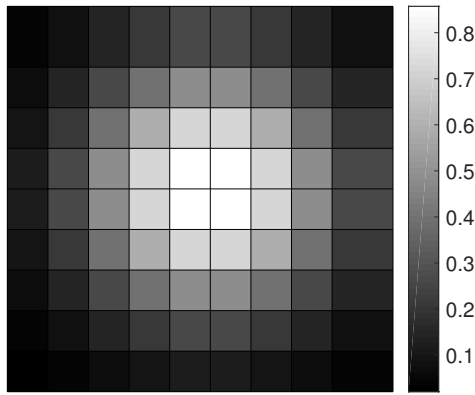

(a) DI $10 \times 10$

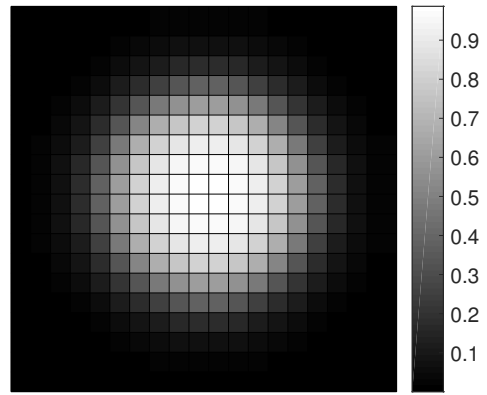

(b) DI $20 \times 20$

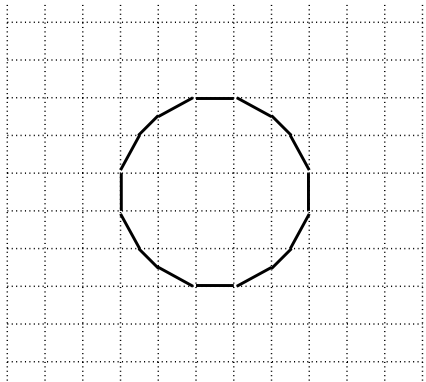

(c) VOF $10 \times 10$

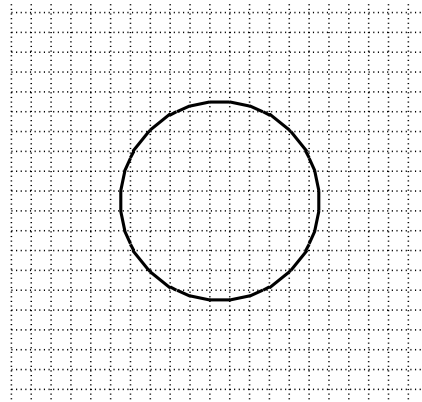

(d) VOF $20 \times 20$

Figure 1: A drop with $R = 0.25$ placed in a $1 \times 1$ domain for DI and VOF at two different resolutions

where the surface tension force per volume is

$$\vec{T} = \sigma \kappa \vec{n} \delta_\Sigma, \quad (4)$$

in which $\delta_\Sigma$ is a numerical Dirac delta function. The convective flux in Equation 3 is computed in a geometric manner consistent with the geometric operator used for volume fraction advection. A balanced force discretization of surface tension forces is used to allow for pressure gradients to balance surface tension forces at equilibrium and reduce spurious currents. Curvature values and the normal vector are computed using embedded height functions (EHF) and utilized in computing the surface tension force per volume, $\vec{T}$, which is implemented as a continuum-surface-force (CSF) as shown in Equation 4. In addition, density and viscosity fields are coupled to Equation 2 via

$$\rho = (\rho_l - \rho_g)f + \rho_g, \quad (5)$$

and

$$\mu = (\mu_l - \mu_g)f + \mu_g. \quad (6)$$

The flow is incompressible,

$$\nabla \cdot \vec{u} = 0, \quad (7)$$

and an adaptation of the fractional-step time integration method of [56] was used for second order time integration and enforcement of Equation 7. It is noteworthy that this novel discretization of the one-fluid formulation discretely conserves mass, and in the absence of viscous and capillary effects, momentum.

2.2. Diffuse Interface Model

Similar to [26], [27] and [38], the phase field chosen for our diffuse interface method has a hyperbolic tangent shape given by:

$$\phi = \frac{1}{2}[1 + \tanh(\frac{s}{2\epsilon})], \quad (8)$$

where $s$ is the signed distance from the interface and $\epsilon$ is a smallness parameter, setting the interface thickness. Moreover, $\phi$ represents the phase field. Note that the hyperbolic tangent interface profile chosen here is the same as the equilibrium profile given by the Cahn-Hilliard equations [35, 33]. However, as shown by [38], the same equilibrium solution can be captured by a second order evolution equation. For an incompressible flow, this equation is given by

$$\frac{\partial \phi}{\partial t} + \nabla \cdot (\vec{u}\phi) = \nabla \cdot \left[\gamma \left(\epsilon \nabla \phi - \phi(1-\phi)\frac{\nabla \phi}{|\nabla \phi|}\right)\right]. \quad (9)$$

We note that the conservative form of this equation naturally ameliorates the problem of mass loss in two-phase flows as compared to level set methods. In other words, the total amount of $\phi$ is conserved globally. The two flux terms on the right hand side are numerical contrivances responsible for preserving the shape of the interface. The second term sharpens the interface while the first term opposes this action by diffusing the phase field, giving a transition region of fixed width. Both terms are multiplied by a common factor related to the maximum velocity in the domain to ensure the interface is held together even in the harshest strain conditions [38],[27], [39]. Both of these constants are set based on findings in [39] such that the discrete solution to the phase field remains bounded and conservative throughout the simulation while taking advantage of second order central finite differences. For simulations presented in this work, we adopt $\epsilon = \Delta x$ and $\gamma = |\vec{u}|_{max}$ where $|\vec{u}|_{max}$ is the magnitude of the maximum velocity in the domain.

In a two-phase flow calculation, Equation 9 must be solved coupled with the Navier-Stokes equation to obtain velocity fields. Surface tension force is included in the Navier-Stokes equations in the form of a continuum surface force [42, 26]. For this, curvature values must be calculated accurately by taking the divergence of the normal vector field. Moreover, the density and viscosity fields are linearly related to the phase field. Thus for an incompressible flow

$$\nabla \cdot \vec{u} = 0, \quad (10)$$

and the Navier-Stokes equations can be written as

$$\frac{\partial \vec{u}}{\partial t} + \nabla \cdot (\vec{u} \otimes \vec{u}) = \frac{1}{\rho} \left\{ -\nabla P + \nabla \cdot [\mu(\nabla \vec{u} + \nabla^T \vec{u})] + \rho \vec{g} + \sigma \kappa \nabla \phi \right\}, \tag{11}$$

where the normal vector ($\vec{n}$) is simply

$$\vec{n} = \frac{\nabla \phi}{|\nabla \phi|}, \tag{12}$$

curvature ($\kappa$) is computed as

$$\kappa = \nabla \cdot \vec{n}, \tag{13}$$

density is linear with respect to $\phi$

$$\rho = (\rho_l - \rho_g)\phi + \rho_g, \tag{14}$$

and similarly for viscosity we have

$$\mu = (\mu_l - \mu_g)\phi + \mu_g. \tag{15}$$

$\rho_l$, $\rho_g$ and $\mu_l$, $\mu_g$ are the densities and viscosities of the two fluids, respectively. Additional implementation details include:

- A staggered grid is used where $P$, $\phi$, $\rho$ and $\mu$ are stored at cell centers. Velocity values are stored at the center of their corresponding cell faces. This is a standard approach that results in desirable conservation properties when central difference schemes are used [57].

- We use second order central differences in space to have mass and momentum conservation (in the absence of surface tension forces) in addition to boundedness of $\phi$. As explained in [39] by choosing the appropriate RHS parameters($\epsilon$ and $\gamma$) for Equation 9, desirable conservation properties can be retained while upwinding schemes or complicated methods for the advection of the phase field are avoided.

- A second order three stage Runge-Kutta stepping scheme adopted from [58] is used for time stepping. This time stepping scheme only requires one Poisson solve per time step.

- The convective term in the momentum equation is in divergence form.

- Since there are no overshoots or undershoots in the phase field equation, no mass-redistribution is required.

2.3. Convergence of Diffuse Interface Methods

When performing convergence studies on diffuse interface methods, the behavior of two sources of error should be taken into consideration [33, 39]. First, there are discretization errors between the numerical solution and exact solution of Equations 9,10,11,13,14,15 which we will refer to as $E_\Delta$. To be clear, this error is because of numerical truncations in approximating the continuous diffuse equations. Second, errors are incurred by approximating a sharp interface with a smoother transitional zone which we denote as $E_\epsilon$. In order to make a diffuse interface simulation converge to sharp interface dynamics one needs to reduce the thickness of the interface ($\epsilon$) and the mesh spacing ($\Delta x$) during convergence studies. Moreover, as explained in [33] and [39], optimal convergence is obtained when both errors scale the same way. The dependence of $E_\Delta$ and $E_\epsilon$ on $\epsilon$, $\gamma$ and $\Delta x$ is however problem dependent. In [39], these errors were studied for the very simple problem of advection of a 1D droplet with a prescribed velocity field. It was found that for that problem, optimal convergence was obtained when $\epsilon/\Delta x$ was kept fixed while the mesh was refined, resulting in a fixed number of mesh points across the interface for different resolutions. Unless explicitly mentioned, we will perform mesh refinement in this manner for the canonical test cases presented herein.

*2.4. Cost Comparison*

An objective and accurate comparison of different numerical methods is a delicate task. There are many challenges in comparing the cost of two codes that consist of multiple subroutines and employ different discretizations and frameworks. To ensure that a fair comparison between the two solvers is performed we have taken the following steps:

- For all the canonical problems (all cases except the droplet impact problem), the simulations were performed on one single processor on the same computing machine.

- While the cost of the DI solver depends almost entirely on the size of the problem (not the density ratio or dynamics), the routines employed in the VOF simulations loop over the interfacial cells, making the run times problem dependent and mainly a function of the number of interfacial cells. Therefore, for each specific problem, we exactly match the input parameters for the two solvers (e.g. densities, surface tension, viscosities and initial conditions), and document the cost for the specific physical problem. This is in contrast to reporting cost based on scaling relations from generic simulations.

- In this work, by cost we are referring to total run time for performing a simulation. In some figures, cost per time-step, which is the run time for one time-step is compared for the different methods.

- In all of the cost figures the relative cost of simulating each case with respect to the coarsest DI calculation (cheapest simulation) is plotted.

- For the droplet impact problem, we use the same clusters to perform the parallel simulations. Overall cost is presented in terms of relative CPU-Hours of computation.

- Generally speaking, the computational cost of an unstructured flow solver is larger than that of an equivalent structured flow solver because the memory layouts of structured flow solvers can be leveraged by the machine (e.g. reduce cache misses). However, the arithmetic intensity of the geometric VOF scheme of Ivey and Moin [13] exceeds that of the momentum solve and curvature computation (the embedded height function scheme of Ivey and Moin [13] reduces to the traditional height function method on uniform hexahedral meshes). The geometric advection method is Lagrangian in nature and has similar costs on both structured and unstructured hexahedral meshes. Therefore, any computational savings provided by a structured memory layout is expected to be negligible compared to the geometric VOF method. As such, the unstructured nature of the VOF solver does not invalidate findings described herein.

- For Cartesian grids, the computational cost of performing VOF simulations in this work is dominated by the computational geometric operations in the advection step. These operations are performed for interfacial cells and computational cost scales with the number of such cells. On the other hand, the cost of the diffuse interface method is dominated by the Poisson solve required to find pressure, and we observe this cost to scale with the total number of cells in the domain. The relative cost of the two methods is problem dependent. For an $N \times N$ two-dimensional domain, if the interface remains local and limited to a narrow region of the domain, the cost of performing one time-step of VOF and DI simulations can be approximated as $C_{VOF}N$ and $C_{DI}N^2$ respectively, where $C_{VOF}$ and $C_{DI}$ are constants and $C_{VOF} \gg C_{DI}$. If the interfacial zones are distributed throughout the domain, like in droplet laden turbulence, bubbly flows or turbulent breaking waves, the cost of one time-step of VOF would then be approximated by $C'_{VOF}N^2$ where again $C'_{VOF} \gg C_{DI}$. This has two implications. First, for cases where the interface is limited to a specific portion of the domain, the relative cost of VOF compared to DI shrinks as the mesh is refined. Second, for cases where the interface is spread around the whole domain, diffuse interface is significantly cheaper regardless of the resolution.

- Although the Poisson solvers used in the two codes are not identical, for the same number of unknowns, the two Poisson solvers give similar run times. Hence, it is safe to assume that the cost comparisons reported in this work are independent of the specific choice of Poisson solvers. It is worth noting however, that typically solving the Poisson system subsumes the majority of the cost of the DI, while only taking a small portion (problem-dependent but generally < 25%) of the cost of the VOF solver [53].

## 3. Results

In this section we consider multiple two-dimensional (2D) canonical problems in which interface transport is coupled with the Navier-Stokes equations. For each test case, we present results from DI and VOF simulations. Many numerical tests that are decoupled from the Navier-Stokes equations are presented in [55, 13] and [21] to verify the accuracy of the numerical methods used for the advection and normal vector/curvature calculation steps of the VOF method chosen for this comparison. Furthermore, the two-phase discretization has been verified using canonical two-phase flows [59]. Similar verification tests can be found in the appendix for the DI method under study herein and in [39]. We have excluded these tests from our comparison section since we intend to make conclusions for realistic applications where the momentum equation and two-phase solver are coupled and computational cost may be a factor which needs to be considered.

We present simulation results from several test cases that evaluate different aspects of the two solvers we have chosen to compare. Section 3.1 uses the problem of advecting a very dense drop to assess the solvers' robustness and ability to handle large density ratios. Next, we proceed to simulations of a more complicated flow, with Rayleigh-Taylor (RT) instability in Section 3.2 providing a measure for comparing the accuracy of interface and momentum transport via the two solvers. This case is unique in the sense that interfacial regions are not confined in space and grow in time, allowing us to study how that affects the cost levels of each solver. Surface tension is zero in the RT simulations, decoupling the accuracy of momentum transport from surface tension force implementation. Moreover, this case has similarities with turbulent two-phase flows as we will explain in Section 3.2. In Section 3.3, spurious currents are reported. We compare the accuracy of the surface tension force implementations via examination of these artificial currents, which are especially detrimental in simulations of surface tension dominant flows. In Sections 3.4 and 3.5, the numerical interplay of inertia and surface tension forces is studied in the context of droplet oscillations and damping surface waves, respectively. Finally, the realistic problem of droplet-pool impact is studied in Section 3.6. This test involves interactions between inertia, surface tension forces and lubrication flow. Moreover, the complex topologies that arise in these simulations provide another opportunity to compare the competence of the solvers.

### 3.1. High Density Ratio Advecting Drop

This test case is adopted from [60], in which a drop of fluid 1 with diameter $D = 0.2$ is advected in the $x$-direction with initial velocity $U = 1$ in a periodic 1×1 box of initially stationary fluid 2. The spatial resolution is chosen to be 128×128 (corresponding to about 25 mesh points across the drop diameter) and no surface tension or viscous forces are present. We consider very high density ratios up to $\rho_1/\rho_2 = 10^7$ to test the robustness of the solvers. There is no exact boundary for the drop in the DI simulations, and the density is very high within the transition zone. In this case, the velocity in the domain is initialized using a hyperbolic tangent profile where the $U = 0.5$ contour lies on points where the density is not very large, say $10^3 \rho_2$. The drop is examined once after traveling a distance equal to one nominal diameter and again after traveling a distance of five nominal diameters (see Figure 2). A CFL number of 0.25 is chosen for these simulations. Theoretically, due to the high density ratio the drop should not "feel" the presence of the surrounding phase and should not deform. The diffuse interface is found to be robust for simulations performed at density ratios up to $\rho_1/\rho_2 = 10^7$. For the VOF method, the simulations are stable for density ratios up to $\rho_1/\rho_2 = 10^5$. For density ratios above that, we resort to lower order advection schemes to keep the simulation running stably. Nevertheless, compared to the DI results, the profile shapes obtained from geometric VOF seem to be closer to the ideal drop shape. This can be attributed to the consistent geometric scheme used for computing convective fluxes in the VOF solver, as explained in [53]. A similar correction can be implemented for the DI momentum transport but is not within the scope of this work.

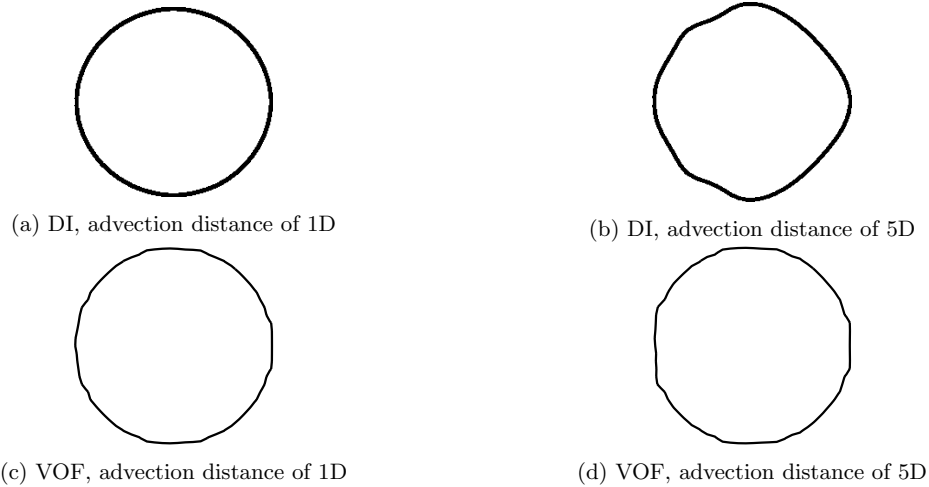

(a) DI, advection distance of 1D    (b) DI, advection distance of 5D

(c) VOF, advection distance of 1D   (d) VOF, advection distance of 5D

Figure 2: Interface profiles for the dense drop advection problem at density ratio of $10^7$ on a $128 \times 128$ grid with around 25 cells across the drop diameter, using DI and VOF with lower order advection operators

The main conclusion from this test case is that both chosen two-phase solvers are very robust in the face of large density ratios.

*3.2. Rayleigh-Taylor Instability*

As we proceed to gradually add more physical effects into the canonical test cases, we study the classical problem of the Rayleigh-Taylor (RT) instability similar to [24]. In this study, no capillary effects exist, so the curvature calculation and surface tension force implementation do not affect the results of this section. There is a dense fluid with $\rho_1 = 1.225$ and $\mu_1 = 0.00313$ sitting on a light fluid with $\rho_2 = 0.1694$ and $\mu_2 = 0.00313$. The interface between the two phases is at $y = 0$ in a $[0, 1] \times [-2, 2]$ box with slip boundary conditions on all sides. Gravity is set to $g = 9.81$ and the interface is initially perturbed with a cosine wave with amplitude 0.05. As the instability grows, the heavier fluid penetrates into the lighter fluid and structures such as the ones shown in Figure 3 emerge. Four different resolutions($64 \times 256, 128 \times 512, 256 \times 1024, 512 \times 2048$) were examined for either of the two-phase methods and Figure 3 shows the interface location at four different times(0.6, 0.7, 0.8, 0.9) long after the linear growth regime of the instability has passed. The VOF simulations show convergence between interfacial profiles at $256 \times 1024$ and $512 \times 2048$, thus the results from $512 \times 2048$ VOF are chosen as reference solutions and plotted in black in the background of Figure 3. On top of the background reference solution in Figure 3, at different resolutions(increasing from left to right) and different times(increasing from top to bottom), interfacial profiles from DI and VOF are plotted on the left (in blue) and right (in red) side respectively. It is clear from this figure that at the same numerical resolution, VOF outperforms DI in capturing the interface location. Apparently however, DI can reproduce similar quality as of VoF when it uses twice the resolution. Note that surface tension is zero in these simulations, so curvature and surface tension force accuracy do not affect the results of this comparison. In other words, this test case assesses the accuracy of phase and momentum advection for DI and VOF, where the coupling is due to the dependence of density and viscosity on the phase (Equations 14 and 15 for DI and 5 and 6 for VOF).

Figure 4a compares the cost of performing the Rayleigh-Taylor simulations on different resolutions using DI and VOF. While DI is found to be an order of magnitude cheaper than VOF at all resolutions, the cost grows as $N_x^2$ for both methods. The cost of DI scales with the number of mesh points in the domain which is expected for a phase field model. The cost of the VOF simulations, however, are mostly due to the advection and surface tension force calculations which scale with the number of interfacial cells. It is apparent from Figure 3 that the number of interfacial cells is growing in time. Therefore, the cost of the VOF calculations increases accordingly as the simulation progresses in time, as shown in Figure 4b. The overall effect is that VOF cost scales with $N_x^2$, which is different from most other simulations we will go through in the remainder of the paper. As mentioned before, we see from Figure 3 that the accuracy of the VOF simulation is comparable to the DI simulation with twice the resolution. On the other hand, from Figure 4a it can be seen that the cost associated with each VOF simulation is comparable to the cost of DI performed at four times that resolution. As a result, we can conclude that at higher numerical resolution but equal cost levels, DI yields more accurate results for this test case. Similar to the RT instability, turbulent two-phase flows also cause interfacial regions to fill the space, resulting in the cost of VOF to scale with the total number of cells. Based on the results in this section, it is reasonable to conclude that for turbulent flows, a higher resolution DI can achieve similar accuracy to a VOF calculation but with less cost.

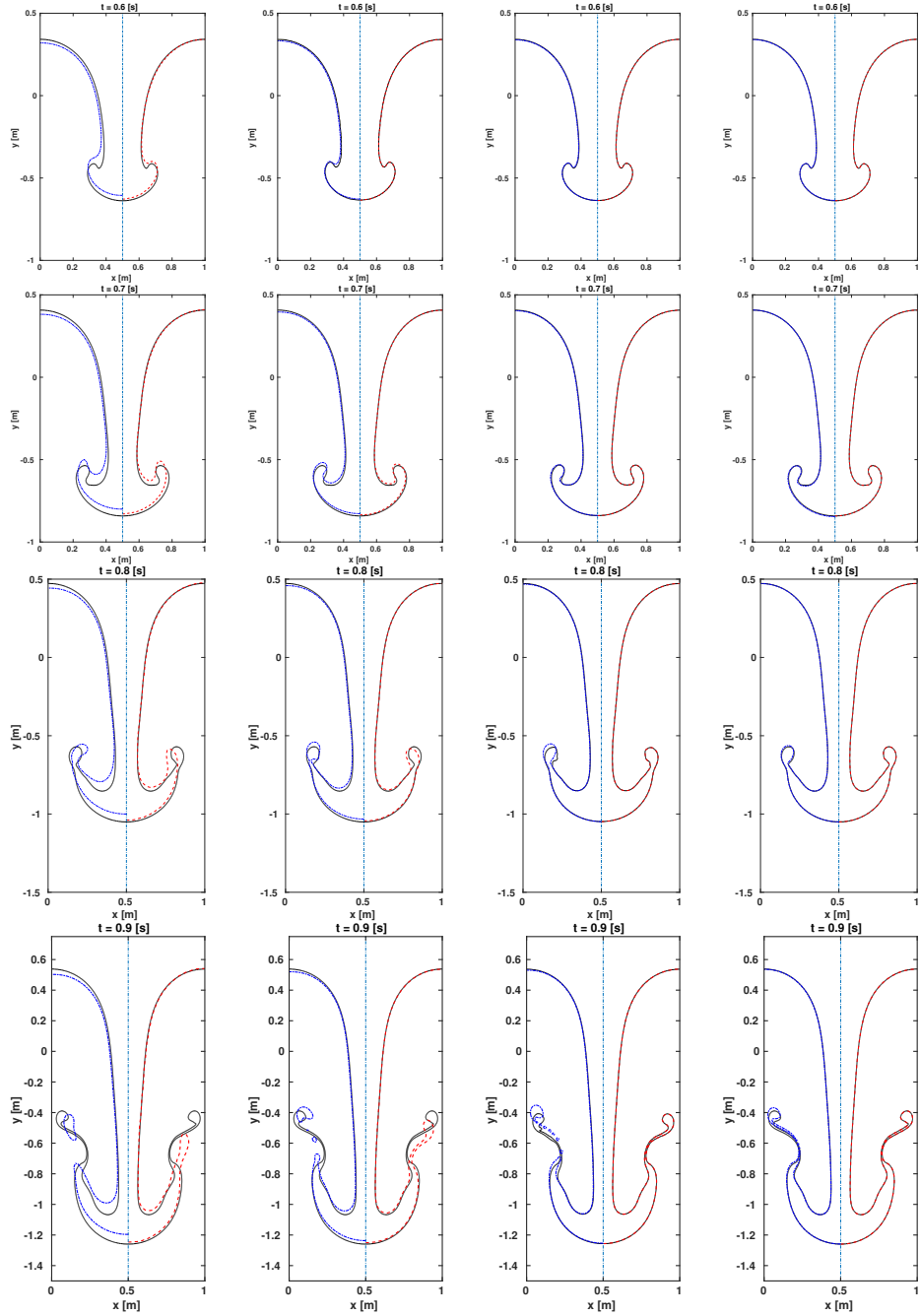

Figure 3: Rayleigh-Taylor instability. Resolution is increased from left ($64 \times 256$) to right ($512 \times 2048$), while physical time increases from top ($t = 0.6$) to bottom ($t = 0.9$). For each panel, blue dashed lines show the interface location predicted by DI on the left and red dashed lines show the interface location from VOF simulations on the right. The reference solution at different times is also plotted in solid black.

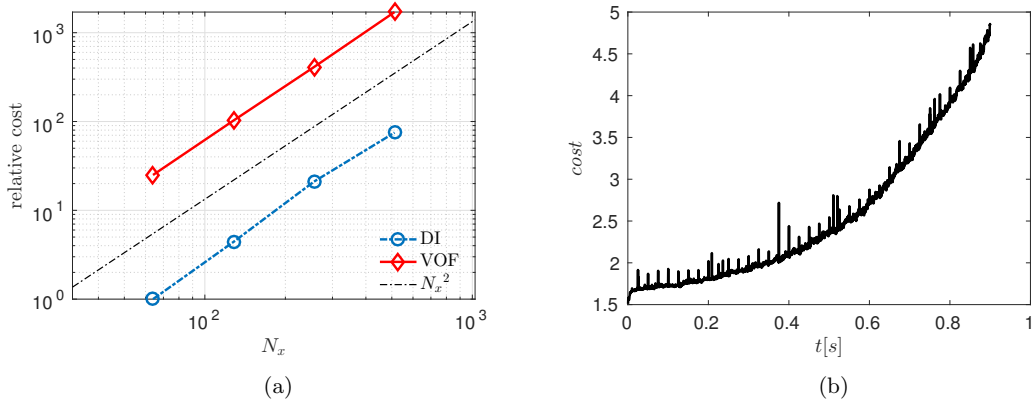

Figure 4: Relative cost of performing Rayleigh-Taylor simulations with DI and VOF compared at different resolutions (a), and cost of performing one time-step of the 64 × 256 VOF simulation plotted against physical time (b)

### 3.3. Spurious Currents in a Stationary Drop

In this test case we have a drop with diameter $D = 0.4$ in a $1 \times 1$ box with slip boundary conditions. The initially stationary drop and surrounding fluid have equal densities and viscosities. The dimensionless parameter characterizing this problem is the Laplace number, $La = \sigma \rho D / \mu^2$. In order to vary $La$, the fluid densities are varied for different tests while everything else including the surface tension ($\sigma = 1$), and viscosity($\mu$=0.1) are kept constant.

If curvatures were computed exactly, the selected DI and VOF solvers would both generate no spurious currents, as they allow for the discrete balance between pressure gradients and surface tension forces, resulting in a stagnant condition as the exact solution to this problem. Therefore, any sustained flow in the numerical solution is due to errors in estimation of curvature from discretized field data, when the surface tension force uses a balanced scheme [61], as done in both methods. We will examine the magnitude of the largest velocity in the domain for $La = 12000$ at non dimensional time $t^* = \sigma t/(\mu D) = 250$, which is a large enough time for the surface tension and viscous forces to have physically balanced each other and for spurious currents to be fully developed. To obtain optimal convergence using DI, we set $\epsilon = \Delta x^{2/3}/\sqrt[3]{32}$ instead of $\epsilon = \Delta x$ which is used for all other test cases in this work. A grid size of 128×128 is used for the cases shown in Figure 5, where the spurious flow field induced by the DI and VOF methods are compared. For the diffuse interface method(Figure 5a), the spurious currents are minimal at locations of the interface that are either aligned with the grid, or angled at ∼ 45 degrees. In addition, the streamlines close to the interface are mostly in the tangential direction and are mostly due to 8 vortex pairs inside and outside of the drop. On the other hand, for VOF(Figure 5b), the flow seems to be due to 8 vortices which are centered on the interface, resulting in maximal influx at angles that are multiples of 45 degrees. This is an interesting observation regarding the flow patterns of spurious currents. Based on our examinations, this subtle contrast in flow patterns seems to be generalizable to all sharp interface (like VOF) and diffuse interface (like phase field) methods, when surface tension force is represented using a CSF form.

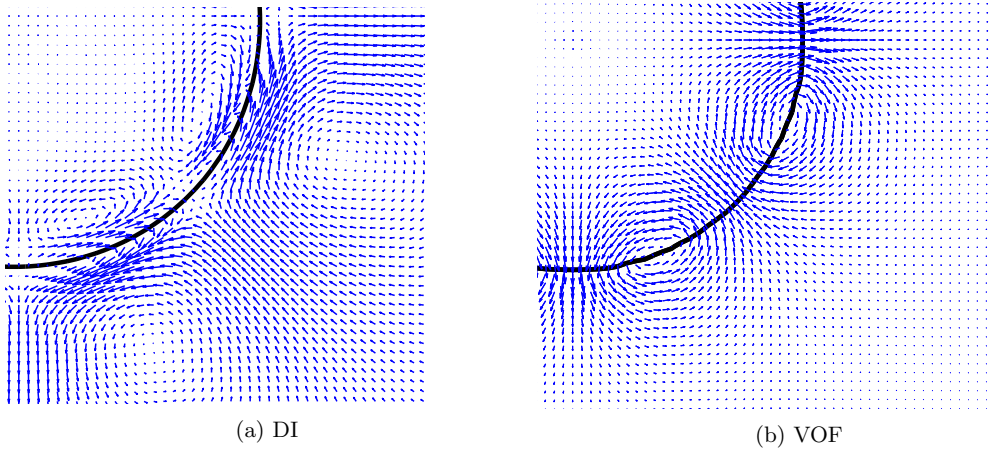

(a) DI  (b) VOF

Figure 5: Zoom-in on spurious currents in and around a stationary drop after equilibrium for the diffuse interface and VOF simulations performed on 128×128 grids

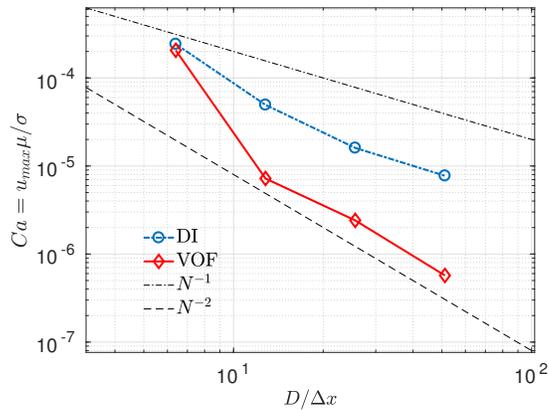

Figure 6: $Ca = u_{max}\mu/\sigma$ associated with spurious flow for the stationary drop problem at different resolutions

In Figure 6 we have compared the magnitude of $Ca = u_{max}\mu/\sigma$ for the two methods at equilibrium time for different resolutions. It is apparent that while the two methods perform similarly at the coarser levels ($\sim$ 6 mesh points across the drop) at higher resolutions VOF produces about an order of magnitude smaller artificial currents compared to the diffuse interface. Moreover, VOF seems to have about second order convergence, while DI has close to first order convergence rate with mesh refinement. These results are in accordance with the predictions of [61], who showed that the magnitude of the maximum spurious current in the domain should have the same convergence rate as the curvature evaluation as one refines the mesh.

We have also compared the accuracy of the two methods against cost for these simulations. In Figure 7a, $Ca = u_{max}\mu/\sigma$ is plotted versus cost of performing one time-step using either method. However, due to its finer resolution at a comparable cost, DI requires finer time-step and thus it is more reasonable to compare the total cost for simulations performed over a fixed physical time ($t^* = 250$). This comparison is plotted in Figure 7b. However, it is also important to consider cost per time-step because in a realistic simulation, the time-step restriction may indeed be imposed by stiffer physics or processes that are happening elsewhere in the domain.

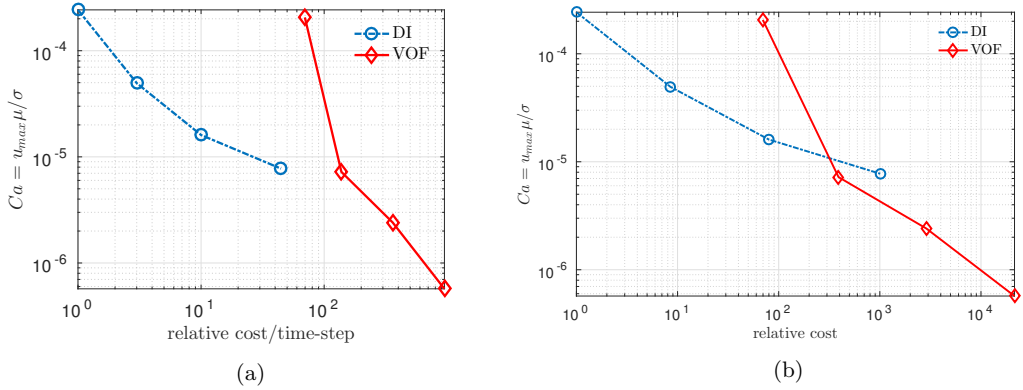

Figure 7: $Ca = u_{max}\mu/\sigma$ in the stationary drop problem at equilibrium plotted against the cost per time step (a) and total cost (b) for DI and VOF

From Figures 7a and 7b, it is apparent that if one desires very high accuracy with minimal tolerance for spurious currents, VOF would be the method that provides lower $Ca$ at a low cost. Nevertheless, in most applications spurious current is an issue at low resolutions (e.g. poorly resolved drops in a liquid atomization process [62] or droplet-laden turbulence [63]), where a refined DI is the better option.

We have also provided comparisons of cost of performing each simulation at different resolutions in Figures 8a and 8b. In Figure 8a the relative cost of performing one time step of either method is plotted. Diffuse interface is consistently one to two orders of magnitude cheaper than VOF at the same resolution. An important distinction between the cost of VOF and DI is highlighted by the scalings of the cost of the two methods. As we previously explained, most of the cost associated with VOF is due to geometric operations for advection and curvature calculation which are performed at cells containing the interface. The number of cells containing the interface scales with $N$, and as a result the whole cost of computation per time-step for VOF scales with $N^\alpha$, where $1 < \alpha < 2$. On the other hand, as explained before, DI is blind to the intricacies of the interface and the cost of all operations (including the sparse matrix Poisson solve step) scales with the number of mesh points, $N^2$. The total cost of reaching $t^* = 250$ is shown in figure 8b. Since the capillary time-step is used here the cost of each time-step is multiplied by a factor of $N^{3/2}$, thus while the ratio of the costs is preserved from Figure 8a to Figure 8b, the total cost evidently scales with $N^\beta$ and $N^{3.5}$ for VOF and DI respectively, where $2.5 < \beta < 3.5$.

3.4. Oscillating drop

In this simulation, a fluid 1 drop of radius $R_0 = 2$ is perturbed with a mode $n = 2$ perturbation and amplitude $A_0 = 0.01R_0$ in a box of size 20×20 filled with fluid 2. Slip boundary conditions are applied on all walls. The physical parameters of the problem are given by $\sigma = 1$, $\rho_1 = 1$, $\rho_2 = 0.01$, $\mu_1 = 1$ and $\mu_2 = 0.0001$. The initial perturbation activates the surface tension forces that causes the drop to oscillate. During these oscillations, the extra energy goes back and forth between surface energy and kinetic energy. The theoretical oscillation frequency for columns in the linear regime was found by [64] to be

$$\omega^2 = \frac{n(n^2 - 1)\sigma}{(\rho_1 + \rho_2)R_0^3}. \tag{16}$$

In our simulations we can measure the period of oscillations, $T_{calc}$, and compare it against the theoretical prediction. Specifically the error of calculated period is reported as

$$E_T = \left|\frac{\omega T_{calc}}{2\pi} - 1\right|. \tag{17}$$

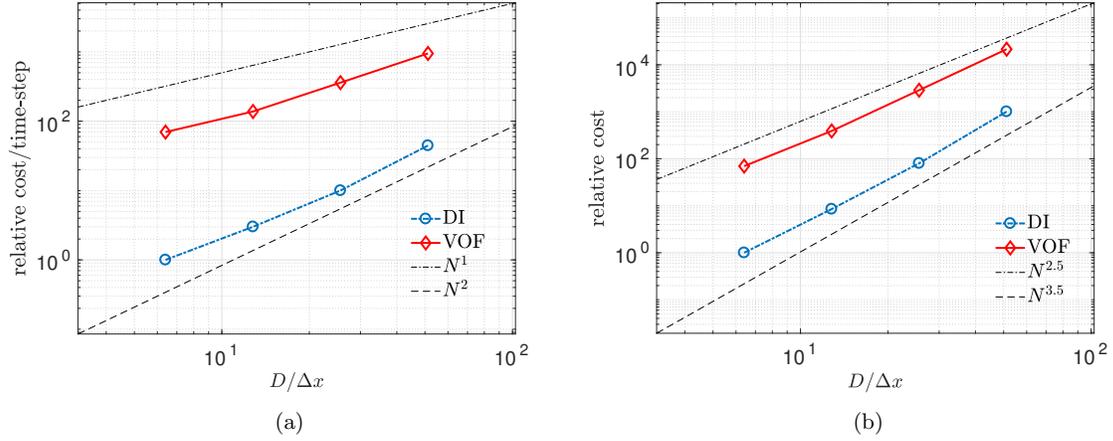

Figure 8: Comparison of cost per time-step (a) and total cost (b) of DI and VOF for the spurious current problem

The error in the calculated period of oscillations from the DI and VOF methods can be seen in Figure 9. By examining this error we assess the accuracy of surface tension force calculations in the two different methods. Similar to previous cases, VOF outperforms DI at the same resolution for this test case. When comparing the period error values from these methods against cost, the conclusion is quite different. This comparison is illustrated in Figure 10. The cost per time-step and total cost are used in this figure to demonstrate that the two methods are performing at comparable levels for this test case.

We provide the relative cost per time-step and relative total cost of DI and VOF in Figure 11. The cost is again around an order of magnitude less for DI. The logic used to interpret Figures 8a and 8b applies here as well. Hence, cost per time-step of DI and VOF scale with $N^2$ and $N^\alpha$ respectively, where $1 < \alpha < 2$. This naturally leads to the total cost scalings of $N^{3.5}$ and $N^\beta$ for DI and VOF respectively, where $2.5 < \beta < 3.5$.

### 3.5. Standing Wave

As the final canonical test case, we look at interfacial oscillations for a standing wave. This problem assesses the accuracy of the two methods at capturing the oscillatory interplay between inertia and surface tension in the presence of viscous dissipation. The mechanism underlying the oscillations is similar to the oscillating drop problem studies in Section 3.4. Initially a single small amplitude wave with wavelength $\lambda = 2\pi$ is placed between two immiscible fluids in a $[0, 2\pi] \times [0, 2\pi]$ domain. The initial perturbation amplitude is $A_0 = 0.01\lambda$ and $y_0 = \pi$. The initial conditions for the wave position are given by

$$h_{wave}(x, t=0) = y - y_0 + A_0 cos(x - \Delta x/2). \tag{18}$$

Boundary conditions are periodic in the x direction and slip boundary conditions hold for the top and bottom walls. If the two fluids have equal kinematic viscosity $\nu$, then the amplitude of the wave will be given by([65]):

$$A_{ex} = \frac{4(1-4\beta)\nu^2}{8(1-4\beta)\nu^2 + \omega_0^2} A_0 erfc\sqrt{\nu t} + \sum_{i=1}^{4} \frac{z_i}{Z_i}(\frac{\omega_0^2 A_0}{z_i^2 - \nu}) exp[(z_i^2 - \nu)t] erfc(z_i\sqrt{t}), \tag{19}$$

where $z_i$ are the roots of

$$z^4 - 4\beta\sqrt{\nu}z^3 + 2(1-6\beta)\nu z^2 + 4(1-3\beta)\nu^{3/2}z + (1-4\beta)\nu^2 + \omega_0^2 = 0. \tag{20}$$

For the test case simulated, $\sigma = 2$ $\rho_1 = \rho_2 = 1$, $\nu = 0.064720863$ and the time step was $dt = 0.003$. The amplitude of the standing wave measured at $x = \Delta x/2$ versus time using DI and VOF is plotted for different resolutions alongside the exact solution in Figures 12a and 12b respectively. It is clear that the sharp interface calculations from VOF are more accurate at capturing the correct location of the interface.

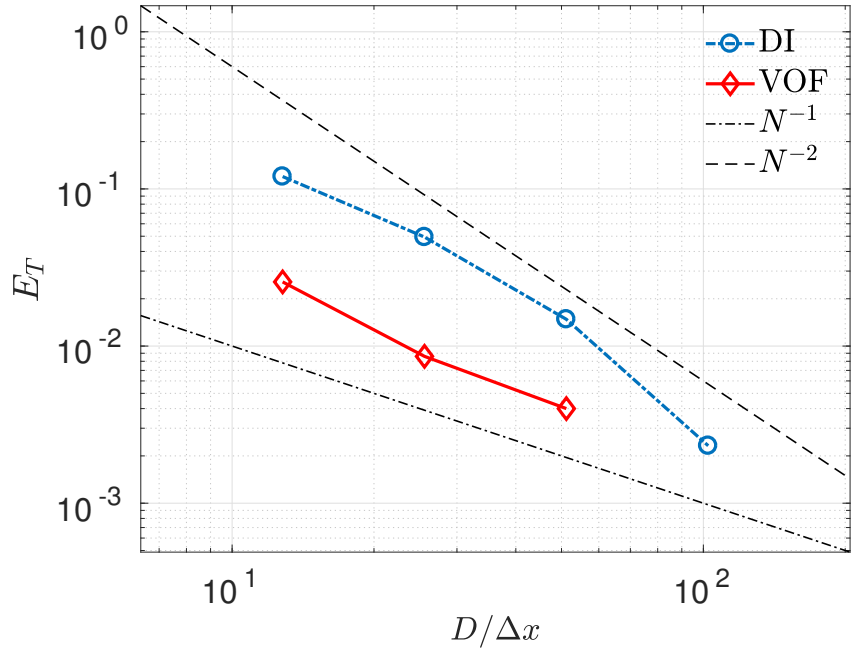

Figure 9: Period error in the oscillating drop problem for different methods and resolutions

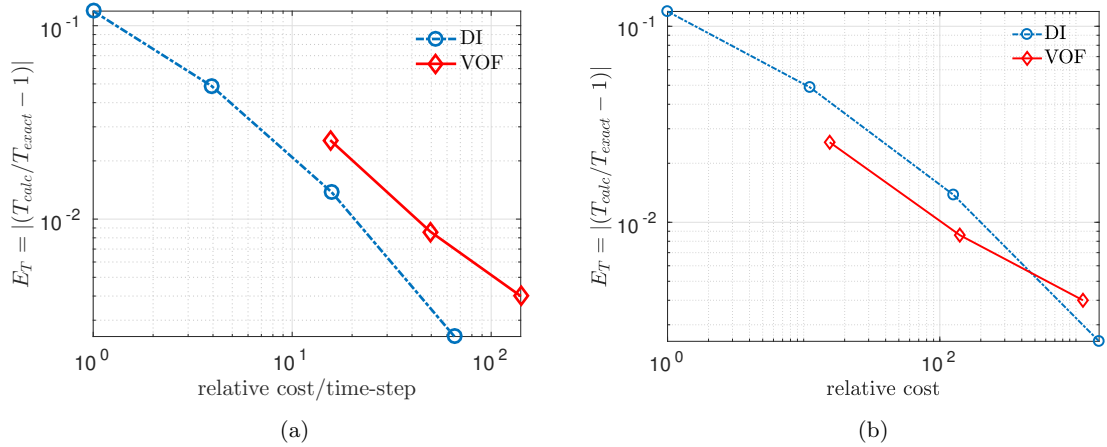

Figure 10: Period Error in the oscillating drop problem for different methods plotted against computational cost per time-step (a) and total computational cost (b)

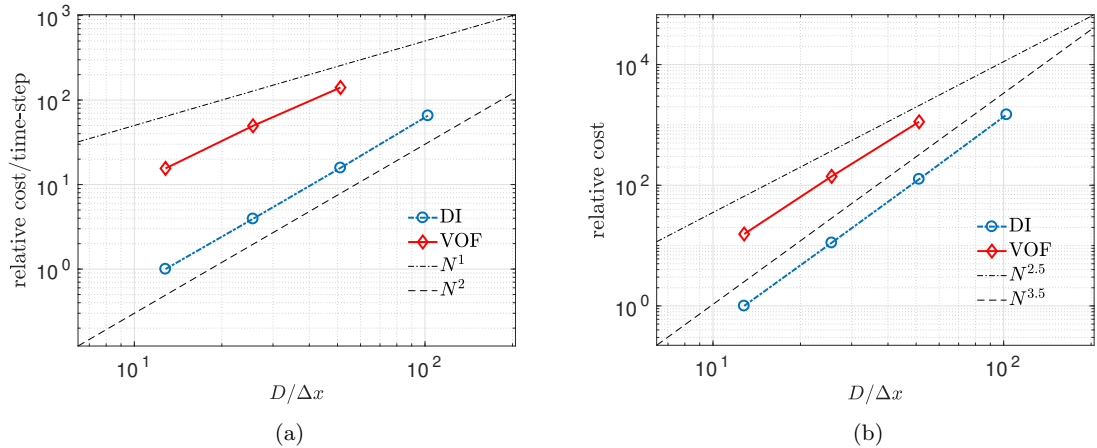

Figure 11: Comparison of cost per time-step (a) and total cost (b) of DI and VOF for the oscillating drop period calculation problem

Using the theoretical amplitude evolution equation (Equation 19) a better comparison of the performance of the two methods can be obtained from examining the root-mean-squared (r.m.s.) error of the simulation predictions. This comparison is given in Figure 13a . While VOF errors for this test case are about an order of magnitude smaller than DI for the same resolution, both methods show first order convergence. It is important to note that the initial amplitude of the wave is small and comparable to the thickness of the interface for the DI simulations. This will lead to errors of the order of the thickness of the interface ($\epsilon$) for the location of the contour $\phi = 0.5$, explaining the first order trend in the errors. Moreover, since our measure for comparison in this test case is the evolution of the small amplitude standing wave, we believe that part of the failure of DI relative to VOF is because of the need for secondary interpolation for finding the exact location of the interface, a step unnecessary for sharp interface models such as VOF.

Once again, we compare the error of the methods against computational cost. Figure 13b shows that VOF is more cost-efficient when very high accuracy in determining the location of the interface is sought. On the other hand, in lieu of a low resolution VOF, DI with higher resolution provides higher accuracy while maintaining the same cost.

In Figure 14, we compare the relative cost of performing the simulations using DI and VOF at different resolutions. Diffuse interface seems to be between one and two orders of magnitude cheaper at the same resolution. However for this topologically simple problem , since the number of interfacial cells clearly scales with $N$, and the majority of the work for the VOF method is done on these cells, the cost seems to be proportional to $N$, as opposed to DI which computes the evolution of $\phi$ in all cells in the domain, resulting in a cost (for Poisson and the rest of computations) scaling with $N^2$.

## 3.6. Droplet Impact

We briefly compare the performance of DI and VOF on a complex problem relevant to air entrainment during droplet-pool impact events. When a water droplet with a $\mathcal{O}(1mm)$ diameter impacts a deep pool of water with normal velocity of $\mathcal{O}(1m/s)$, a thin gas film between the drop and pool cushions the impact. This air layer becomes entrapped between the two liquid bodies after they contact at a distance from the center, and then retracts to leave a large bubble behind. Figure 15a shows a schematic showing the thin film with respect to the drop and pool. While it is clear from the experimental images in [66] that the stages after the thin film is entrapped require 3D simulations, the events leading to the formation of the thin gas film can be well approximated with 2D simulations [67]. Moreover, although in these experiments undulations appear on the edge of the retracting thin film, these azimuthal instabilities do not lead to pinch-off of smaller bubbles. As such a 2D simulation of the full problem is also expected to result in one large bubble after the thin film has retracted.

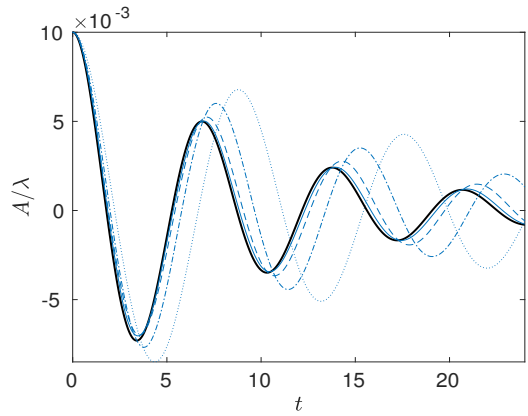

(a) DI

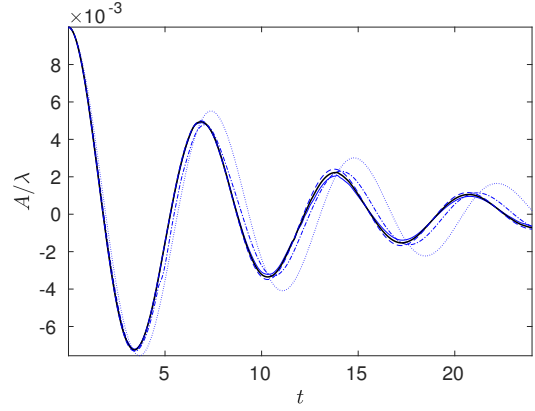

(b) VOF

Figure 12: Amplitude of the wave at $x = \Delta x/2$ from DI and VOF simulations are plotted against time for different resolutions alongside the theoretical prediction. In both figures, the theoretical curve is plotted in solid black, while the numerical results from $16 \times 16$, $32 \times 32$, $64 \times 64$ and $128 \times 128$ simulations are plotted using dotted, dashed-dotted, dashed and solid blue lines respectively.

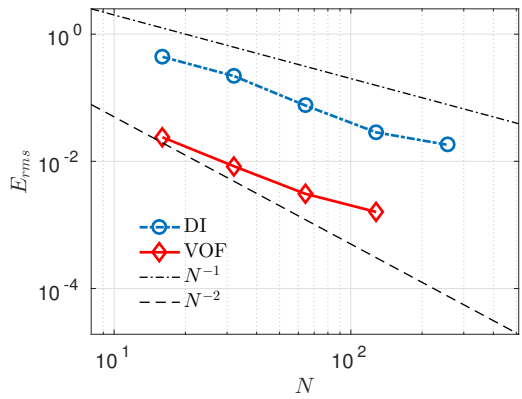

(a)

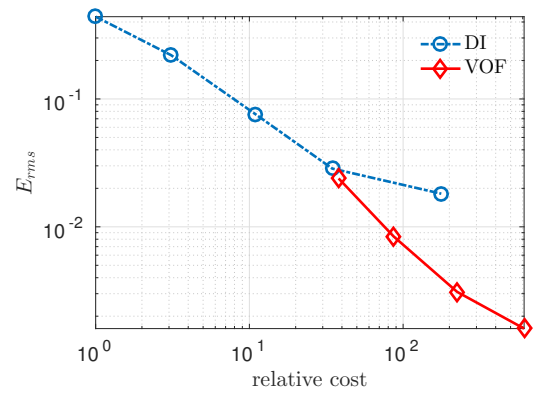

(b)

Figure 13: Standing wave r.m.s error values for different methods plotted against resolution (a), and standing wave r.m.s error values for different methods plotted against computational cost (b)

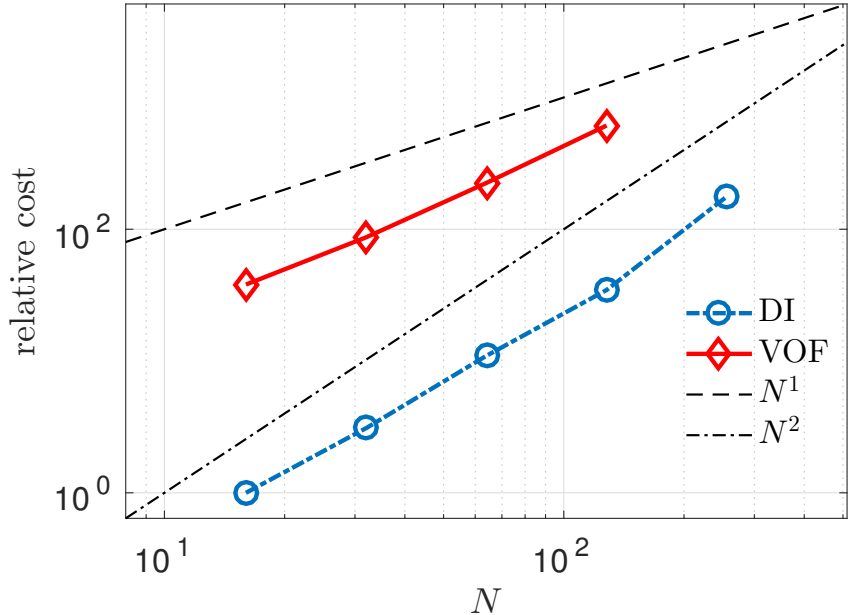

Figure 14: Cost comparison of DI and VOF for the standing wave problem

Boundary element methods (BEM) coupled with lubrication flow in the gas are usually employed for simulating this phenomenon [67, 68]. Such methods are preferred since they do not require resolving the thickness of the very thin gas film. Nevertheless, in this section we compare the performance of VOF and DI at capturing the outcome of the drop-pool impact problem using highly resolved 2D two-phase simulations.

We simulate a water-air system with impact velocity $U = 1m/s$ and droplet diameter $D = 3.4mm$ on stretched meshes concentrated around the region of impact. Due to the resolution requirements of this problem, we perform these simulations utilizing parallel computing. In Figure 15b, the thin gas film is shown just prior to the contact of the liquid bodies, depicting how the simulations converge to a solution that is also close to the predictions of a BEM simulation similar to [67]. The mean film thickness of the film is also in agreement with experimental measurements by [69]. After the drop and pool contact however, some smaller bubbles are observed in the simulations. These bubbles are unphysical, as their size shrinks almost proportional with the mesh, and also as the experiments show, only one large bubble remains after retraction of the thin gas film. These numerical bubbles can be seen in Figure 15c, which shows a zoom-in on the contact zone of a 6144 × 6144 DI simulation. The diameter of the large bubble and the numerical bubbles generated from VOF and DI are sorted and plotted in Figure 16. Indeed, the size of these artificial bubbles shrink linearly with mesh refinement as shown in Figure 16.

Through rigorous comparisons we find that both methods converge on the geometry of the film and the diameter of the large physical bubble resulting from it's retraction, but VOF requires less resolution as one would expect. Figure 16 shows how results from a DI simulation on a 6144 × 6144 stretched mesh capture features with the same length scales as a 2048 × 2048 stretched mesh VOF simulation of this problem. Curiously however, the VOF simulation (with 3 times mesh points in both directions) requires about 5 times more computational cost than the DI simulation. It's worth a reminder that in the VOF method

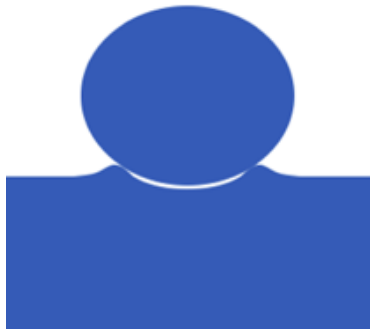
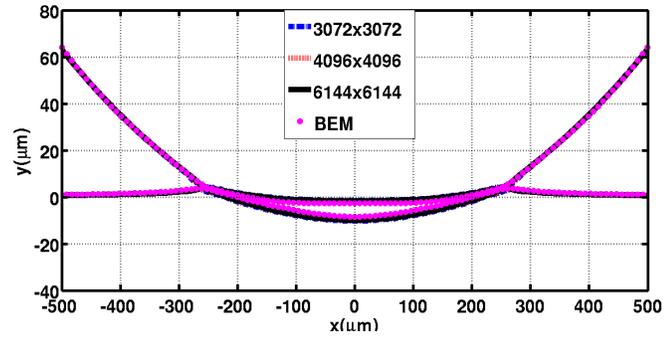

(a)
(b)

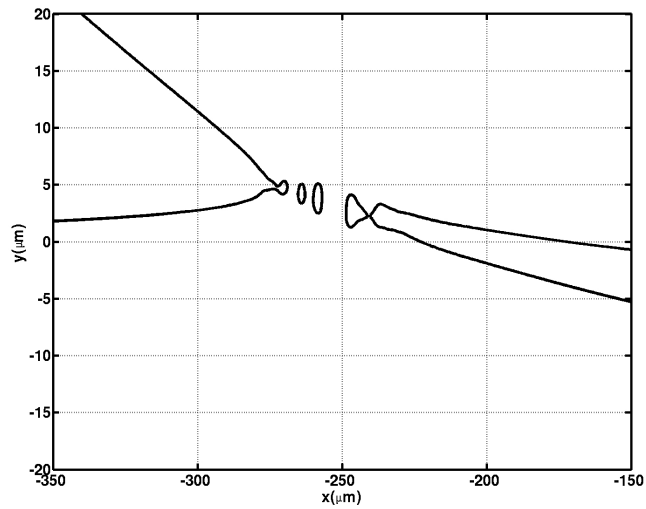

(c)

Figure 15: Schematic of how a thin gas film is entrapped during the drop-pool impact event (a). Entrapped thin film at contact time from DI simulations at three different resolutions and a BEM simulation for a $D = 3.4mm$ water drop impacting a deep pool with $U = 1m/s$(b). A zoomed-in snapshot of the interface and numerical air bubbles around the region where the drop and pool first contact for a $6144 \times 6144$ DI simulation of a $D = 3.4mm$ water drop impacting the pool with $U = 1m/s$ (c)

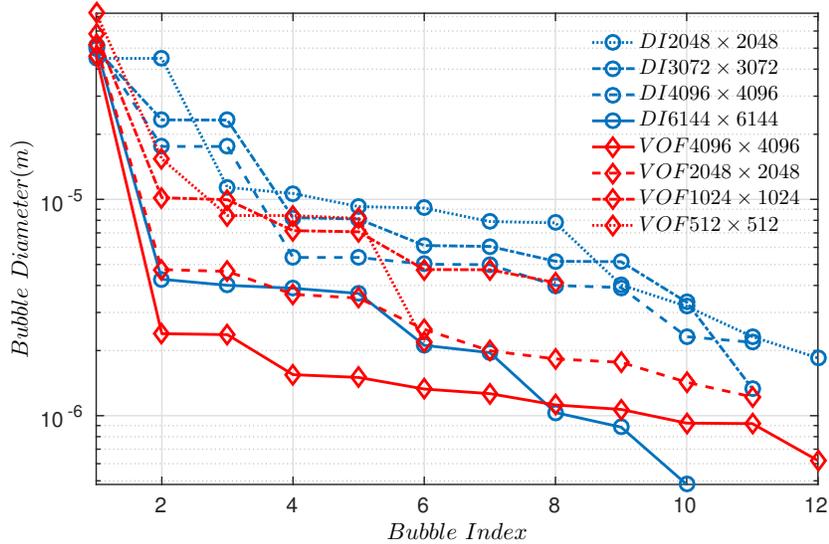

Figure 16: Sorted bubble diameters for VOF and DI, demonstrating convergence on the size of the large bubble while numerical bubbles shrink with mesh refinement

studied here, the operations deployed to find the volume fractions and their normal vector are performed in regions containing the interface. As a result, when a highly stretched mesh is used to concentrate most mesh points around the interface, the cost almost scales with the number of total mesh points. Moreover, in such large parallel computations, load balancing is much more difficult for VOF compared to DI which is load-balanced in nature.

## 4. Conclusions

In this paper we compared two fundamentally different yet promising numerical methods from the class of VOF and DI methods. While both of these methods have attracted plenty of attention from researchers, an unbiased comparison of these methods' capabilities was warranted. Several test cases were chosen to assess the relative performance and computational cost of these methods at solving the interface evolution equation coupled to the Navier-Stokes equation.

It was observed that for most of the test cases VOF was more accurate than DI at the same resolution. In terms of convergence behavior, both methods were convergent on all test cases with comparable convergence rates. However, the computational cost of DI was typically an order of magnitude less than VOF at the same resolution. We generally observed that while maintaining accuracy, one could save a factor of about 2-3 in resolution by switching from DI to VOF. However, more often than not, the cost of these two comparably accurate simulations would be similar or even cheaper in the case of DI.

For most test cases, DI with more resolution was found to be a better option than coarse or under-resolved VOF simulations. At high resolutions VOF yielded higher accuracy at lower cost compared to DI. This was especially found to be true for problems where the interfacial cells were confined in space. For such flows, if the space has $\lambda$ dimensions and N cells per direction, the cost of VOF and DI scale with $N^{\lambda-1}$ and $N^{\lambda}$ respectively, giving VOF an advantage at higher $N$ values. We verily observed this for some of our small canonical problems; however, large realistic problems requiring parallel computing incur load balancing issues for the geometric VOF, while posing no extra challenge to the scalable DI. Alternatively if the interface is not confined to a particular region of space, as is the case for most high $Re$ two-phase flows, high resolution would be needed at all points in space. For a flow in $\lambda$ dimensions, this results in the VOF

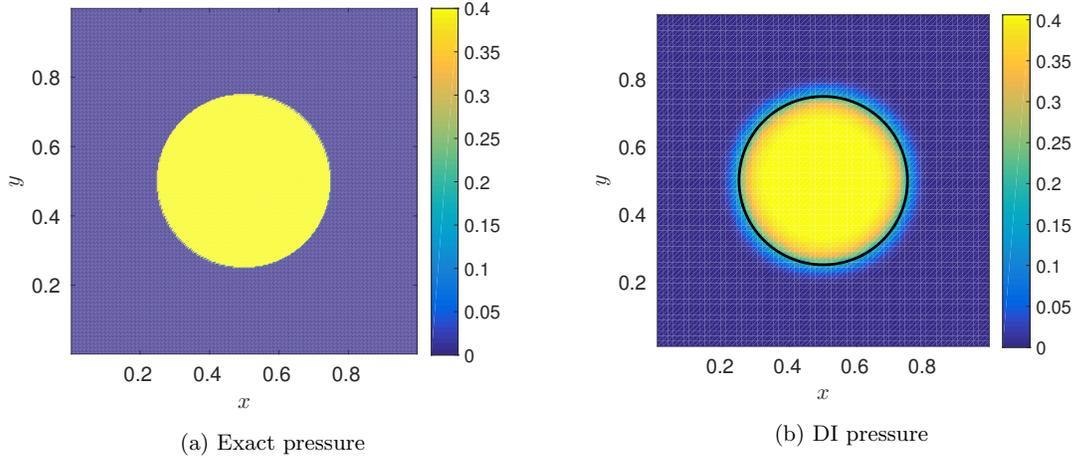

(a) Exact pressure

(b) DI pressure

Figure 17: Pressure field from diffuse interface compared to the theoretical value of pressure for a sharp interface drop. In addition, the interface location is plotted in black in part (b)

cost to scale with $N^\lambda$, similar to DI. As such, compared to VOF, the low computational cost of DI would allow for cheaper DI calculations at higher accuracy and cost levels for these problems. Based on these discussions and our results, it seems likely that in both types of flows, the accuracy of DI accompanied by its scalability and low cost allow for DI to outperform VOF in many practical simulations.

## Appendix A  Diffuse Interface Tests

### A.1  Curvature Accuracy Test

From Equations 11 and 13 we know that the surface tension force takes the form of $\sigma \kappa \nabla \phi$, where curvature is obtained from the normal vector using $\kappa = \nabla \cdot \vec{n}$. Calculating curvature error is different between diffuse interface methods and sharp interface methods. While in sharp interface methods curvature calculation is only undertaken at the interface, in diffuse interface methods curvature must be accurately computed in a transition layer between the two phases. Strictly speaking, the interface exists everywhere in the domain; thus, a global indicator would be more suitable to study the surface tension force calculation accuracy. In the absence of the inertial and viscous terms in Equation 11, pressure gradients must balance the surface tension forces. Hence, the value of the pressure drop across an interface with known curvature is a practical indicator of the accuracy of surface tension forces. Specifically, we will utilize a test case where a circle of radius $R = 0.25$ is placed in the center of a $1 \times 1$ box and the pressure computed from DI will be compared to the theoretical pressure field(Figure 17). Note that theoretically, the pressure jump at the drop interface(corresponding to $\phi = 0.5$) is given by $\Delta P = \sigma \kappa / R$. If we neglect the error in solving the linear Poisson system for pressure, the error in pressure is solely due to the errors committed during surface tension force calculation in DI. On a Cartesian mesh, this error can be calculated via

$$E_P = \sum_{i=1}^{N_x} \sum_{j=1}^{N_y} \left| P(i,j) - HS(R - r(i,j)) \frac{\sigma}{R} \right| \Delta x \Delta y, \qquad (21)$$

where $HS(x)$ is the heavy side step function, and $r$ is the distance of a point from the center of the domain. In Figure 18, the pressure error, $E_P$ is plotted against resolution. These calculations were performed by selecting $\epsilon = \Delta x = \Delta y$ in all simulations. Clearly, the error in the pressure calculation decreases linearly with mesh refinement.

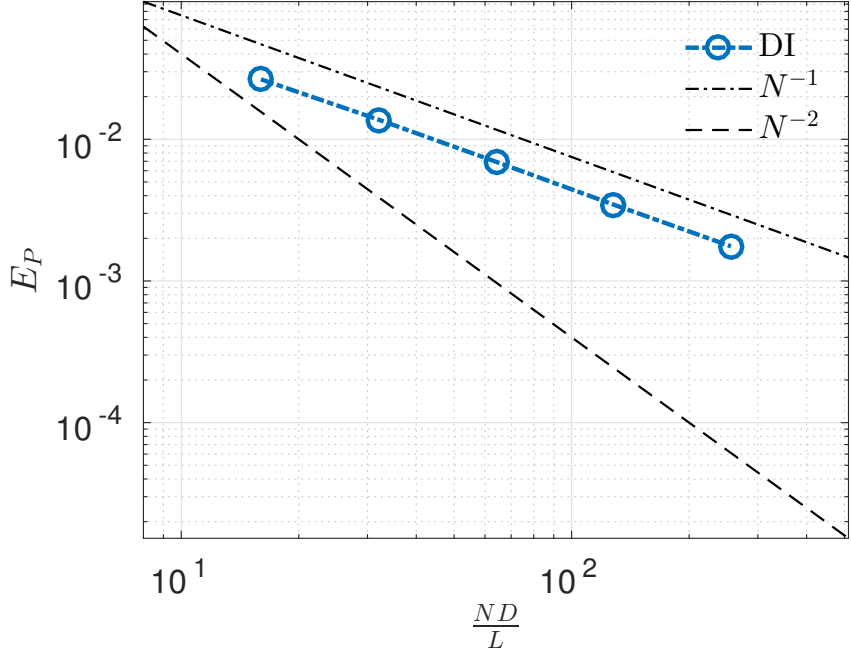

Figure 18: Convergence study on the accuracy of pressure/surface tension force calculation

*A.2   Advection Test*

A standard test for two-phase flow advection schemes is the 2D drop in shear flow test case. A column of radius $R_0 = 0.15$ and center (0.5; 0.75) is placed inside a unit sized box. The velocity field is given by the stream function,

$$\Psi(x,t) = \frac{1}{\pi} sin^2(\pi x) sin^2(\pi y) cos(\frac{\pi t}{T}) \qquad (22)$$

, and $T = 8$. The velocity field specified for this test case dictates the initial circle to be retained at the end of simulation time. Hence, an insightful check is to compare the final captured interface with the exact solution. For these test cases, a fixed time step of $dt = 0.5\Delta x$ was chosen. In Figure 19, the state of the drop at $t = T/2$ can be seen for different resolution simulations. As Figure 19 shows, the drop gets stretched in the vortex field and artificial drops appear at low resolutions. In Figure 20, the drops interfacial profiles at the end of the simulation(t=T), which should analytically be the same as initial conditions is plotted against the initial drop for different resolutions.

If we define the final error in the shape of the drop to be defined by:

$$E_{shape} = \sum_{i=1}^{N_x}\sum_{j=1}^{N_y} |HS(\phi(i,j,t=T) - 0.5) - HS(\phi(i,j,t=0) - 0.5)| \Delta x \Delta y \qquad (23)$$

Then, Figure 21 shows how errors are reduced with approximately second order convergence rate.

## 5. Acknowledgements

This work was supported by Office of Naval Research (Grant No. 119675) and NASA (Grant No. 127881).

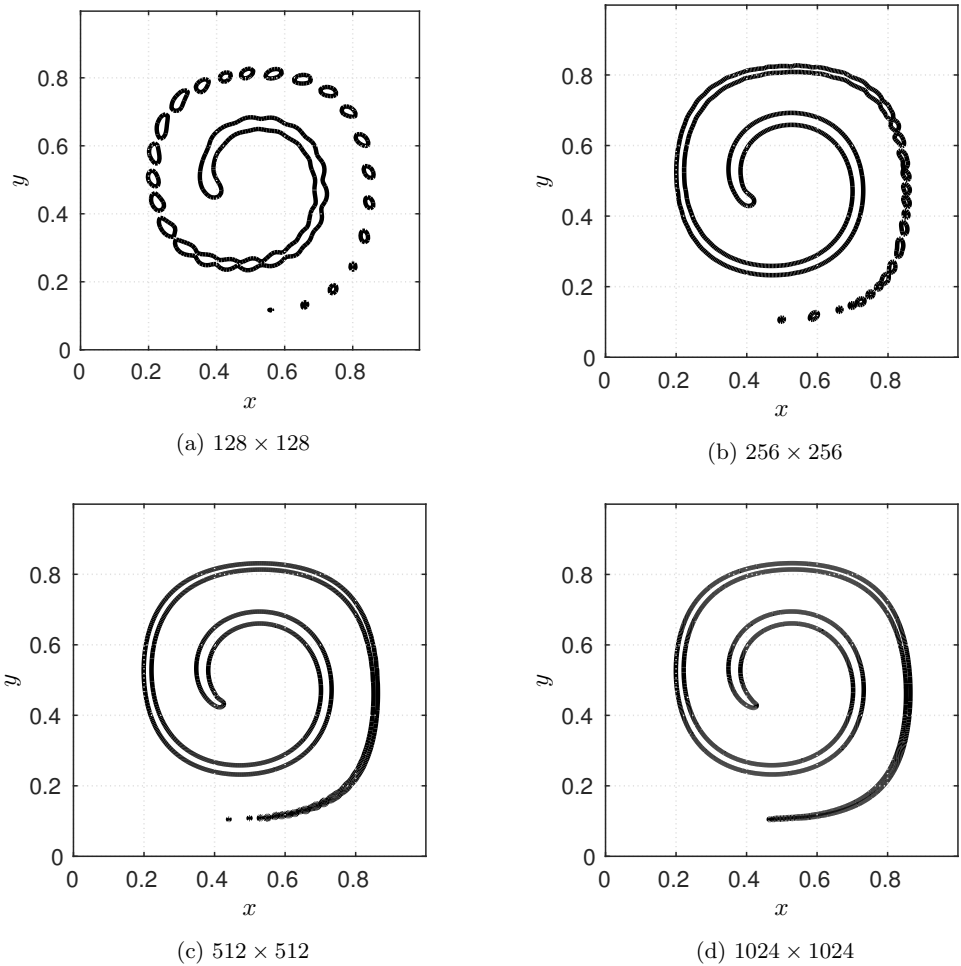

Figure 19: Interfacial profiles at $t = T/2$ for drop in deformation field. Increasing resolution from left to right and top to bottom results in an intact ligament which does not break into drops

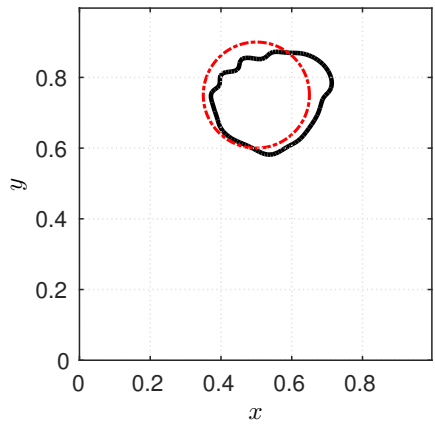
(a) $128 \times 128$

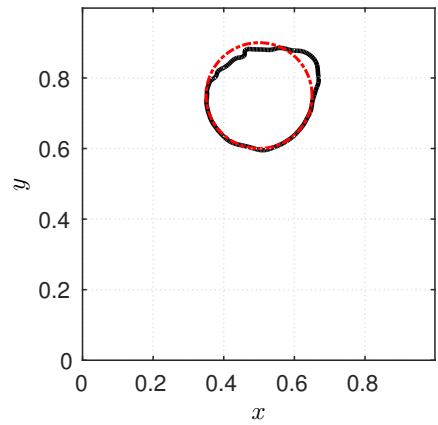
(b) $256 \times 256$

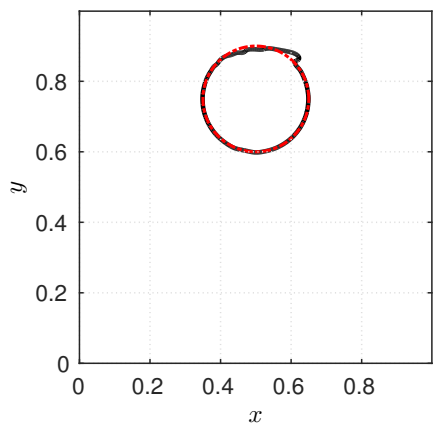
(c) $512 \times 512$

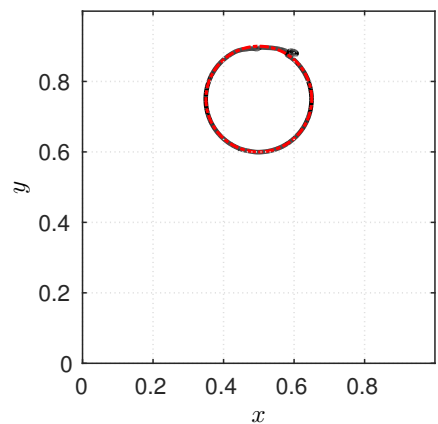
(d) $1024 \times 1024$

Figure 20: Interfacial profiles at $t = T$ for drop in deformation field plotted against initial drop. The final solution and initial drop should lay on top of each other with a perfect two-phase flow numerical method.

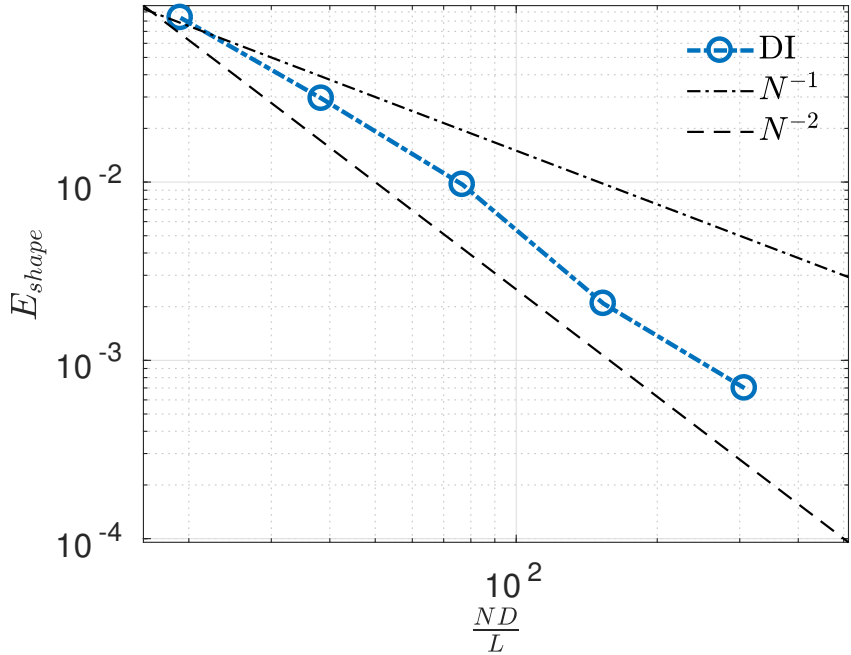

Figure 21: Shape error for the deforming drop test case versus resolution